\documentstyle[prl,aps,multicol,epsfig]{revtex} 
 
\begin{document}                
 
\def\be{\begin{equation}} 
\def\ee{\end{equation}} 
\def\ba{\begin{eqnarray}} 
\def\ea{\end{eqnarray}} 
 
\title{Experimental Test of a Trace Formula   
for a Chaotic Three Dimensional Microwave Cavity} 
\author{C.~Dembowski,$^1$ B.~Dietz,$^1$ H.-D.~Gr\"af,$^1$ A.~Heine,$^1$  
T.~Papenbrock,$^2$ A.~Richter,$^1$ and C.~Richter$^1$} 
\address{$^1$ Institut f\"ur Kernphysik, Technische Universit\"at Darmstadt,  
D-64289 Darmstadt, Germany} 
\address{$^2$ Physics Division,  
Oak Ridge National Laboratory, Oak Ridge, TN 37831-6373, USA} 
\date{\today} 
\maketitle 
\begin{abstract} 
 
We have measured resonance spectra in a superconducting microwave 
cavity with the shape of a three-dimensional generalized Bunimovich stadium  
billiard and analyzed their spectral 
fluctuation properties. The experimental length spectrum  
exhibits 
contributions from periodic orbits of non-generic modes and from  
unstable periodic orbits  
of the underlying classical system.  It  
is well reproduced by 
our theoretical calculations based on the trace formula derived by 
Balian and Duplantier for chaotic electromagnetic cavities. 
 
\end{abstract} 
 
  
\begin{multicols}{2} 
\narrowtext

In the last few decades, billiard systems have provided a most 
appropriate model for the understanding and discussion of classical 
chaotic Hamiltonian systems \cite{SinBerBun} and their 
quantum counterparts \cite{BohigHeller}. 
In two dimensions the Schr\"odinger equation for quantum billiards 
coincides with the Helmholtz equation for microwave cavities of 
corresponding shape.  
This correspondence is the basis for experimental 
studies of quantum manifestations of classical chaos. Early 
experiments focused on spectral fluctuation properties in chaotic 
billiards \cite{stoeckmannbuch,R1}.   
About 30 years ago 
Gutzwiller established a direct relationship between the 
oscillating part of the density of states and the properties of the 
classical periodic orbits by means of a trace formula 
\cite{Gutz}.  Gutzwiller's trace formula and its extension 
to systems with mixed classical dynamics \cite{Ullmo} were verified 
experimentally using flat microwave cavities \cite{R1,mix}.  
Three-dimensional chaotic billiard systems have been scarcely studied 
theoretically \cite{PS,Prosen}.  
There is no analogy between quantum billiards and electromagnetic 
cavities in three dimensions. Still, the latter are of great interest 
for the study of wave dynamical phenomena in chaotic systems.   
Spectral properties of the vectorial Helmholtz equation have recently 
been studied theoretically for integrable systems \cite{Frank} and 
experimentally for mixed and chaotic systems 
\cite{Weav,Wirzba,SinaiExp}. 
 
\begin{figure} 
\centerline{\epsfxsize=4.35cm \epsfbox{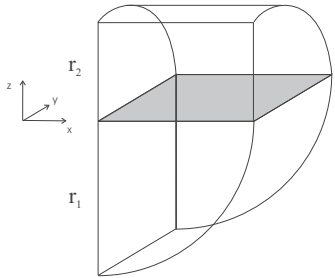}}\vspace*{2.5ex} 
\caption{Desymmetrized version of the three-dimensional 
generalized stadium billiard. The plane $z=0$ is shaded.}                   
\label{fig1} 
\end{figure}

\begin{figure} 
\centerline{\epsfxsize=6.5cm \epsfbox{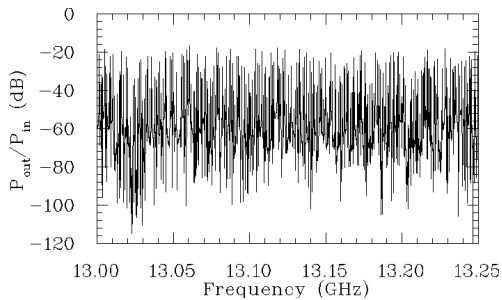}}\vspace*{2.5ex} 
\caption{Typical transmission spectrum for the frequency range 
13-13.25~GHz, where the resonances are still well separated.}                   
\label{fig2} 
\end{figure}

\begin{figure} 
\centerline{\epsfxsize=6.5cm \epsfbox{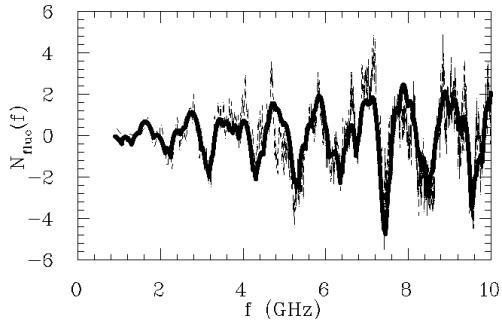}}\vspace*{2.5ex} 
\caption{ 
Fluctuating part of the experimentally obtained staircase function 
(dashed line) compared to the fluctuating part of the staircase 
function for the non-generic modes (full line). We only show that 
range of the spectrum where both curves are distinguishable.} 
\label{fig3} 
\end{figure}

\begin{figure} 
\centerline{\epsfxsize=6.5cm \epsfbox{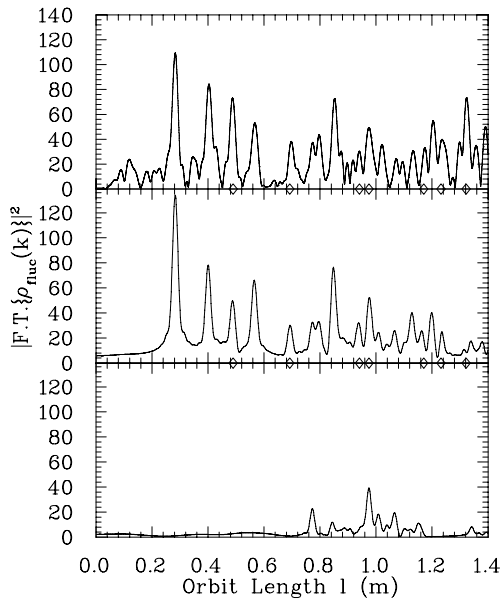}}\vspace*{2.5ex} 
\caption{ 
Length spectra from the experimental data (top) compared to (middle) 
the sum of non-generic modes [from Eq.~(\ref{nongen})] plus the 
unstable periodic orbits [trace formula (\ref{BD}) of Balian and 
Duplantier], and the unstable periodic orbits (bottom). Diamonds mark 
those peaks that are assigned to periodic orbits in the plane $z=0$.} 
\label{fig4} 
\end{figure}

We present the 
first experimental test of the trace formula derived by Balian and 
Duplantier \cite{BaDu} for chaotic electromagnetic resonators. The 
main difficulty to overcome consists in the construction of a microwave 
resonator that is completely chaotic while permitting only a few 
non-generic modes. We meet these requirements by using a 
superconducting resonator with the shape of a desymmetrized 
three-dimensional stadium billiard \cite{Stadium3D}.  
Figure~\ref{fig1} 
shows that this billiard consists of two quarter cylinders with radii  
$r_1$ and $r_2$, resp. 
The electromagnetic resonator has dimensions $r_1=200.0~{\rm mm}, 
\ r_2=141.4~{\rm mm}$ and is made of niobium which becomes 
superconducting at temperatures below 9.2 K. This tremendously 
increases the resolution of the measured spectra due to a quality 
factor of up to $10^7$ compared to $10^3$ in normal conducting 
resonators. The measurements were performed at a temperature of 4.2 K 
for frequencies  $f$ up to $20$~GHz. We show a typical spectrum 
in Fig.~\ref{fig2}.  
The evaluation of four reflection and six transmission  
spectra yielded 18764 resonances. 
According to Weyl`s formula \cite{BaDu,Weylt} for electromagnetic 
cavities the smooth part of the integrated resonance density,  
${N_{\rm smooth}(f)}$, is a polynomial of third order in the 
frequency, where, in contrast to the corresponding quantum case, the 
quadratic term is absent. Its coefficients   
have been obtained by a fit to ${N(f)}$. The  
fluctuating part $N_{\rm fluc}(f)$ is shown in Fig.~\ref{fig3}; it still  
displays smooth  
oscillations, which are due to the non-generic periodic orbits. 
Our cavity (see Fig.~\ref{fig1}) exhibits two types of non-generic orbits.  
First, there are 
two families of three-dimensional, marginally stable ``bouncing ball'' 
orbits with length $2r_1$ and $2r_2$ that evolve parallel to the axis 
of the two cylinders. Second, there are 
trajectories inside the plane $z=0$ that are linearly stable with 
respect to deviations parallel to this plane and unstable with respect 
to perpendicular deviations. While both types of orbits are of measure 
zero, 
they generate smooth oscillations in the staircase function and yield 
dominant peaks in the length spectrum, i.e. the absolute value of the 
Fourier transform of the fluctuating part of the $k$-dependent density of  
states, where $k=2\pi f/c_0$ is the wave number, $c_0$ denotes the 
speed of light.  
Such effects of non-generic periodic orbits 
have been found in various types of billiards \cite{R1,SinaiExp,PP}. 
The experimental length spectrum is shown in 
the top of Fig.~\ref{fig4}. 
To obtain an analytical expression for the staircase function  
of the non-generic modes we combine the semiclassical 
method of Ref.\cite{Wirzba} with the adiabatic method of Ref.\cite{Taylor}. 
The  quantum adiabatic method \cite{Taylor} has been  
applied to the calculation of bouncing ball modes in the two-dimensional  
stadium billiard.  
This method corresponds to a Born-Oppenheimer approximation, where  
the fast coordinate is  
parallel to the classical motion corresponding to the bouncing ball  
orbits, while the slow variable is transversal to it.  
Accordingly, we assume that the modes in the 
$x-$ and $y-$directions are adiabatically decoupled from the modes in 
the $z$-direction and quantize the rectangle with side lengths $l_x$ and 
$l_y$. Then, the $x-$ and $y-$component of the wave vector $\vec k_{\rm ng}$ 
of the non-generic modes are given as $k_{x}={\pi\mu /l_x}$ and  
$k_{y}={\pi\nu /l_y}$ for  
integers $\mu$ and $\nu$, where, according to the geometry of our 
billiard the $z$-dependence of the lengths $l_x$ and $l_y$ is 
\[ \begin{array}{lllrl} 
l_x(z)=r_1, & l_y(z)=\sqrt{r_2^2-z^2} & ~~\mbox{for} & 0 & \le z\le r_2 \\ 
l_y(z)=r_2, & l_x(z)=\sqrt{r_1^2-z^2} & ~~\mbox{for} & -r_1 & \le z < 0. 
\end{array}  
\] 
Combining this adiabatic method with the approach of ref. \cite{Wirzba} we  
express the staircase function 
$N_{\rm ng}(k)$ for the non-generic modes (ng) as $N_{\rm ng}(k)={\rm Tr}  
\Theta(k^2-k_{\rm ng}^2)$, where the trace is over the wave vector   
$\vec{k}_{\rm ng}$ of the non-generic modes    
\be 
N_{\rm ng}(k)=\sum_{\mu,\nu}\int{dz dk_z\over 2\pi}\,  
\Theta\left(k^2-k_x^2-k_y^2-k_z^2\right). 
\label{nongen} 
\ee 
Note that the integration over $z$ is restricted to the interval 
$[-r_1,r_2]$ while the $k_z$-integration is unrestricted. The integers 
$\mu$ and $\nu$ label the modes in $x-$ and $y-$directions, resp. 
Dirichlet ($\mu,\nu>0$) or von Neumann ($\mu,\nu\ge 0$)  
boundary conditions correspond to the magnetic 
and electric non-generic modes, resp.  
 
\begin{figure} 
\centerline{\epsfxsize=6.5cm \epsfbox{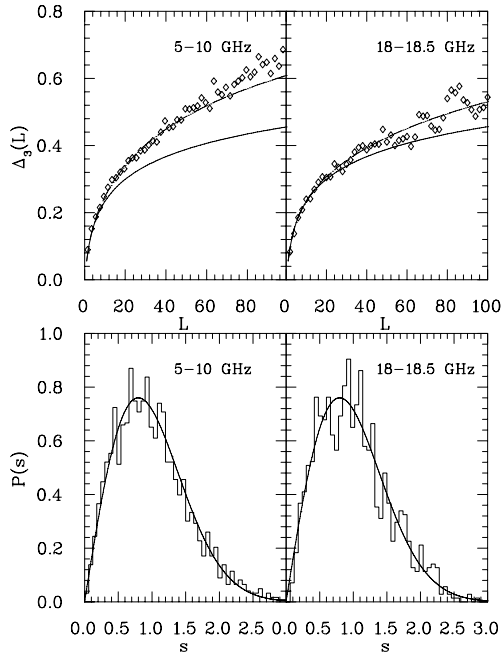}}\vspace*{2.5ex} 
\caption{Left figures: Experimental level-spacing 
distribution (histogram) and $\Delta_3$-statistics (diamonds) for the 
frequency range 5-10~GHz ($\lambda=0.352$) compared to GOE   
(full line) and the predictions from the 
model (see Eq. (\ref{model})) (dashed line). Right figures: Same as left 
figures but for the frequency range 18-18.5~GHz ($\lambda =0.658$).} 
\label{fig5} 
\end{figure}

For the electromagnetic cavity under consideration, the non-generic 
magnetic and electric modes decouple. Thus, the non-generic 
contribution to the staircase function is given by the sum of the 
quantum mechanical expression for Dirichlet and von 
Neumann boundary conditions. 
Figure~\ref{fig3} shows that the smooth oscillations of the 
fluctuating part of the experimental staircase function are well described  
by our expression. Thus, the adiabatic method yields 
a very good approximation for the contributions of the non-generic 
modes to the staircase function.  
In order to study the local fluctuation properties in the resonance 
spectrum, we rescaled the resonances to unit mean spacing and subtracted 
the non-generic contributions from the spectrum. For a 
time-reversal invariant, classically chaotic system the local 
fluctuation properties are expected to coincide with those of random 
matrices from the Gaussian orthogonal ensemble (GOE) 
\cite{Mehta,BGS}. We however find notable deviations and attribute 
them to a partial decoupling between electric and magnetic modes which is 
prominent at low frequencies. Indeed, our spectral statistics are in 
very good agreement with that obtained for random matrices $\hat H$ 
from an ensemble, that models two coupled, chaotic systems 
\cite{GMW,Rosen,Alt}, 
\ba  
\label{model} 
\hat H=\left(\matrix{ \hat H_e &\sqrt{\lambda}Dv_{ij}\cr 
\sqrt{\lambda}Dv_{ji} &\hat H_m\cr}\right).   
\ea  
Here, $\hat H_e$ and $\hat H_m$ are matrices from the GOE with 
dimensions $N_e$ and $N_m$, resp. The coupling matrix elements 
$v_{ij}$ are Gaussian random variables with zero mean and unit 
variance, $D$ is the mean level spacing, and $\lambda$ is the coupling  
strength. The random matrix 
model interpolates between two uncoupled GOE-systems at $\lambda=0$ 
and one GOE at $\lambda=1$. We obtain the best agreement between our 
experimental level statistics and that of random matrices $\hat H$  
as defined in Eq. (\ref{model}) when we treat 
$\lambda$ as a fitting parameter and set $N_e$ ($N_m$) equal to the 
number of eigenvalues of the scalar Helmholtz equation with von 
Neumann (Dirichlet) conditions on the boundary of the billiard. 
This suggests an interpretation of model (\ref{model}) in 
terms of electric and magnetic modes that are coupled with strength 
$\lambda$. Figure~\ref{fig5} shows the level-spacing distribution and 
the $\Delta_3$-statistics for a frequency range of 5-10~GHz ($\lambda =0.352$) 
and 18-18.5~GHz ($\lambda =0.658$). The level-spacing distributions agree well  
with the GOE 
and with those of the model defined by Equation (\ref{model}).  
The $\Delta_3$-statistics coincides   
with that for the random matrix model (\ref{model}) but differs from the 
GOE. Note, however, that the GOE is approached with increasing 
frequency, (2-5~GHz: $\lambda =0.07$, 11.5-13~GHz: $\lambda =0.546$, 15-17~GHz: 
$\lambda =0.633$). 
We furthermore found that $2 \pi \lambda = \Gamma /D$, where 
$1/D$ ist the level density, while $\Gamma$ shows only a weak dependence 
on frequency for the frequency range from 0 to 20 GHz. 
This is similar to a behavior that was observed in studies 
of isospin mixing in nuclei \cite{harricwei86}. 
The partial decoupling of electric and magnetic modes at low 
frequencies was not observed in the three-dimensional Sinai billiard 
\cite{SinaiExp} and seems to be due to the particular geometry of the  
 three-dimensional generalized stadium billiard. Its desymmetrized version 
(see Fig.~\ref{fig1}) corresponds to two quarter  
cylinders, which are rotated by 
$\pi / 2$ with respect to each other. The electric and the magnetic modes 
are completely decoupled in each of these quarter cylinders \cite{Frank}.  
 
Let us finally turn to a semiclassical analysis of the experimental spectrum. 
According to \cite{BaDu} the generic 
contribution to the fluctuating part of the density of states is  
semiclassically given by the periodic orbit sum, i.e. the trace formula 
\be 
\rho_{\rm fluc}(k)=\sum_{p}{2\cos{(\phi_p)}L_p/\pi\over  
|{\rm det}(1-M_p)|^{1/2}} 
\cos{\left(kL_p-{\pi\over 2}\mu_p\right)}. 
\label{BD} 
\ee 
The factor $2\cos{\phi_p}$ stems from the polarization and is 
thus due to the vectorial character of the underlying wave 
equation. It vanishes for orbits with an odd number of 
reflections.  
The remaining quantities in 
Eq. (\ref{BD}) are identical to those appearing in Gutzwiller's 
trace formula \cite{Gutz}. 
We numerically determined the first 381 periodic 
orbits, their lengths $L_p$ up to about 1.5~m, and their stability 
matrices $M_p$ by two different methods. Based on an ansatz for a 
symbolic code (see, e.g., Chap. 8 in ref.~\cite{stoeckmannbuch}) 
we determined all orbits in the full (not the desymmetrized)  
three-dimensional stadium billiard for up to eight reflections off  
the boundary. For more reflections the numerical evaluations  
become too time consuming. The resulting list 
of periodic orbits was checked for completeness and enlarged by a numerical  
search of fixed points in the Poincar{\'e} surface of section $(z=0, 
p_z>0)$. The latter method is blind to some whispering gallery modes.  
As in the  
two-dimensional stadium billiard our device exhibits infinitely many  
whispering gallery modes. The shortest of these orbits 
propagate along the curved edges of the billiard and 
have lengths $2r_1+\pi r_2 \approx 0.843$m, $2r_2+\pi r_1\approx 
0.910$m and $\pi(r_1+r_2)\approx 1.07$m. There are no significant 
peaks associated with these lengths (see Fig.~\ref{fig4}, top).  
Similar cancellation effects have been observed in the two-dimensional stadium 
billiard \cite{2dstad}.  
For the computation of the Maslov indices 
$\mu_p$ we followed \cite{Sieber}. Figure~\ref{fig4} shows a 
comparison between the experimentally obtained length spectrum, the length  
spectrum for the sum of  
non-generic orbits [Eq.~(\ref{nongen})] plus  
unstable periodic orbits [Eq.~(\ref{BD})], and the length spectrum  
of the unstable periodic orbits alone.   
Diamonds mark those peaks of the non-generic contributions, that cannot be 
assigned to bouncing ball modes but to orbits in the plane $z=0$.  
Evidently, the theoretical reconstruction describes the  
experiment rather well. One 
peculiarity are peaks that appear at about $l\approx 0.77~$m and $l\approx 
0.97~$m. While we could not find corresponding orbits inside the 
billiard, there are two orbits in the boundary plane $y=0$, whose 
lengths and stability amplitudes exactly match those peaks. 
Peaks at lengths below $l=0.28~$m,  
which corresponds to the shortest possible periodic orbit, 
are the result of the (unavoidable) experimental inaccuracy. 
The good agreement between the experimental and the theoretical 
length spectrum quickly deteriorates beyond 
$l\approx 1.3~$m. We recall the  
exponential proliferation of long orbits which enter  
the trace formula~(\ref{BD}). It might be that we 
have missed some  periodic orbits in our  
numerical reconstruction of the length spectrum. 
Note that a considerable fraction of the short periodic orbits have 
$|2\cos{\phi_p}|=2$. In a semiclassical picture the electric and 
magnetic modes decouple locally on such an orbit. This finding 
supports our interpretation of the level statistics in terms of a 
partial decoupling between electric and magnetic modes.  
 
In summary, we have investigated wave chaotic phenomena in a 
superconducting three-dimensional microwave resonator. Spectral fluctuations  
on short 
frequency scales agree well with those of random matrices from an 
ensemble modelling two coupled chaotic systems. We interpret this 
as a partial decoupling of the electric and magnetic modes in the 
low-frequency domain. The length spectrum can be understood in terms 
of non-generic modes and in terms of unstable periodic orbits of the 
underlying classical system. 

We acknowledge discussions with B. Eckhardt, T. Guhr, H.~L. Harney,   
T.~H. Seligman, S. Tomsovic, and T. Weiland. T.P. thanks the 
MPI f\"ur 
Kernphysik, Heidelberg, for its hospitality during the initial stages 
of this work. Oak Ridge National Laboratory is managed by UT-Battelle, 
LLC for the U.S. DOE under contract 
DE-AC05-00OR22725. This work was supported by the DFG 
under contract Ri 242/16-1 and -2 and  
by the HMWK within the HWP. C.D., B.D., A.H., 
and T.P. thank CONACYT for support during the workshop {\it Chaos in few 
and many body problems} at CIC.

 
\end{multicols} 
\end{document}